\documentstyle[12pt]{article}

\newcommand{\bmat}{\left(\begin{array}}
\newcommand{\emat}{\end{array}\right)}
\def\NPB#1#2#3{Nucl. Phys. B{#1} (19#2) #3}
\def\PLB#1#2#3{Phys. Lett. B{#1} (19#2) #3}

\def\PRD#1#2#3{Phys. Rev. D{#1} (19#2) #3}
\def\PRL#1#2#3{Phys. Rev. Lett. {#1} (19#2) #3}

\def\Deq#1{\mbox{$D$=#1}}

\def\yzero{\smash{\hbox{$y\kern-4pt\raise1pt\hbox{${}^\circ$}$}}}

\def\g{\gamma}

\def\Om{\Omega}
\def\om{\omega}
\def\th{\theta}

\def\-{\hphantom{-}}
\def\ov{\overline}
\def\s2{\frac{1}{\sqrt2}}

\def\beq{\begin{equation}}
\def\eeq{\end{equation}}
\def\beqa{\begin{eqnarray}}
\def\eeqa{\end{eqnarray}}

\def\Tr{{\rm Tr \,}}
\def\diag{{\rm diag \,}}
\def\IF{\relax{\rm I\kern-.18em F}}
\def\II{\relax{\rm I\kern-.18em I}}
\def\IP{\relax{\rm I\kern-.18em P}}
\def\IC{\relax\hbox{\kern.25em$\inbar\kern-.3em{\rm C}$}}
\def\IR{\relax{\rm I\kern-.18em R}}

\def\Dsl{\,\raise.15ex\hbox{/}\mkern-13.5mu D} 
\def\IZ{Z\kern-.4em  Z}
\def\id{{\rm I}}

\catcode`\@=11
\newdimen\@rotdimen
\newbox\@rotbox

\def\@vspec#1{\special{ps:#1}}
\def\@rotstart#1{\@vspec{gsave currentpoint currentpoint translate
   #1 neg exch neg exch translate}}
\def\@rotfinish{\@vspec{currentpoint grestore moveto}}
\def\@rotr#1{\@rotdimen=\ht#1\advance\@rotdimen by\dp#1%
   \hbox to\@rotdimen{\hskip\ht#1\vbox to\wd#1{\@rotstart{90 rotate}%
   \box#1\vss}\hss}\@rotfinish}
\def\@rotl#1{\@rotdimen=\ht#1\advance\@rotdimen by\dp#1%
   \hbox to\@rotdimen{\vbox to\wd#1{\vskip\wd#1\@rotstart{270 rotate}%
   \box#1\vss}\hss}\@rotfinish}%
\def\@rotu#1{\@rotdimen=\ht#1\advance\@rotdimen by\dp#1%
   \hbox to\wd#1{\hskip\wd#1\vbox to\@rotdimen{\vskip\@rotdimen
   \@rotstart{-1 dup scale}\box#1\vss}\hss}\@rotfinish}%
\def\@rotf#1{\hbox to\wd#1{\hskip\wd#1\@rotstart{-1 1 scale}%
   \box#1\hss}\@rotfinish}%
\def\rotate{\@ifnextchar[{\@rotate}{\@rotate[l]}}
\def\@rotate[#1]#2{\setbox\@rotbox=\hbox{#2}\@nameuse{@rot#1}\@rotbox}

\catcode`\@=12

\topmargin
-1.5cm
\textwidth
15.5cm
\textheight
23.5cm
\oddsidemargin
0.7cm
\evensidemargin
1.2cm

\begin{document}

\makeatletter
\@addtoreset{equation}{section}
\makeatother
\renewcommand{\theequation}{\thesection.\arabic{equation}}
\pagestyle{empty}
\rightline{FTUAM-99/10; IFT-UAM/CSIC-99-12; IASSNS-HEP-99/33}
\rightline{\tt hep-th/9904071}
\vspace{0.5cm}
\begin{center}
\LARGE{
Tadpole versus anomaly cancellation in D=4,6 compact IIB orientifolds  \\[10mm]}
\large{G.~Aldazabal$^{1,2}$,
D.~Badagnani$^2$,
L.~E.~Ib\'a\~nez$^3$ and A.~M.~Uranga$^4$
\\[2mm]}
\small{
$^1$ CNEA, Centro At\'omico Bariloche,\\[-0.3em]
8400 S.C. de Bariloche, and CONICET, Argentina.\\[1mm]
$^2$ Instituto Balseiro, Universidad Nacional de Cuyo,\\[-0.3em]
8400 S.C. de Bariloche, Argentina. \\[1mm]
$^3$ Departamento de F\'{\i}sica Te\'orica C-XI
and Instituto de F\'{\i}sica Te\'orica  C-XVI,\\[-0.3em]
Universidad Aut\'onoma de Madrid,
Cantoblanco, 28049 Madrid, Spain.\\[1mm]
$^4$ Institute for advanced Study, Olden Lane, Princeton NJ 08540, USA.
\\[9mm]}
\small{\bf Abstract} \\[7mm]
\end{center}

\begin{center}
\begin{minipage}[h]{14.0cm}
It is often stated in the literature concerning $D=4,6$ compact Type IIB
orientifolds that tadpole cancellation conditions i) uniquely fix the
gauge group (up to Wilson lines and/or moving of branes) and ii) are
equivalent to gauge anomaly cancellation. We study the relationship
between tadpole and anomaly cancellation conditions and qualify
both statements. In general the tadpole cancellation conditions
imply gauge anomaly cancellation but are stronger than the latter conditions
in $D=4$, $N=1$ orientifolds. We also  find that  tadpole cancellation
conditions in $Z_N$ $D=4,6$ compact orientifolds do not completely fix the
gauge group and we provide new solutions different from those previously
reported in the literature.

\end{minipage}
\end{center}
\newpage
\setcounter{page}{1}
\pagestyle{plain}
\renewcommand{\thefootnote}{\arabic{footnote}}
\setcounter{footnote}{0}

\section{Introduction}

It is a well known fact that anomaly cancellation in $SO(32)$ Type I
string may be understood as a direct consequence of the cancellation of
tadpoles of Ramond-Ramond fields. In fact the implication runs in both
directions and anomaly cancellation implies also tadpole cancellation.
This is not so surprising since in $D=10$ the anomaly cancellation
constraints fix almost uniquely the massless particle content of the
theory.

In Type I vacua in lower dimensions, like in Type IIB $D=4,6$ orientifolds,
tadpole cancellation constraints do also imply anomaly cancellation. An
interesting question arises regarding the extent up to which anomaly
cancellation and tadpole cancellation are still equivalent in lower
dimensions. One of the purposes of this paper is to address this question.
We concentrate in the study of $D=4$, $N=1$ Type IIB compact orientifolds,
although analogous results are obtained in $D=6$. We find that tadpole
cancellation is in general a more stringent constraint than anomaly
cancellation in the case of $D=4$ orientifolds. In order to show this, we
rewrite the anomaly cancellation conditions in terms of traces of Chan-Paton
twist matrices acting on 9-branes and/or 5-branes. In this way, it is
shown that in general only a subset of the complete tadpole cancellation
conditions is recovered.  Thus, there are certain tadpole cancellation
constraints which are actually {\it not} required for anomaly cancellation.
Which those are depends on the structure of the twist group of the
orientifold and also on the simultaneous presence or not of 9-branes and
5-branes\footnote{We are referring here to chiral $D=4$ orientifolds.
Non-chiral ones like the $Z_2\times Z_2$ model of ref.\cite{bl}
obviously do not
get any constraint from anomaly cancellation.}

This procedure allows a broad scan for solutions of tadpole cancellation
conditions in $D=4,6$ $N=1$, Type IIB orientifolds. In particular, we find
that there are certain $D=4$ and $D=6$ orientifolds which admit many more
solutions than previously reported in the literature.

The relationship between tadpole cancellation and anomaly cancellation was
considered in ref.\cite {lr} \footnote{See also \cite{bi1,bi2,pu} for
analogous results in six-dimensional orientifolds.} in the context of
$D=4$, $N=1$ gauge theories on the world-volume of D3-branes sitting at
$Z_N\times Z_M$ singularities \footnote{$N=1$ supersymmetric theories
form D3 branes at orbifold singularities have been studied in \cite{dgm,
ks,lnv,hu,hh}. The inclusion of orientifold projections has been discussed
in \cite{lykken,kak1,kak2,iru}.}. In that reference an equivalence between
tadpole  and anomaly cancellation conditions was found. Those models
differ from the class we consider in several important respects. In
particular they are non-compact models (the six dimensions transverse to
the D3-branes are not compactified) and in addition  they only contain
D3-branes.

In this paper we consider {\it compact} orientifolds and this fact makes
the following important difference. Tadpoles are sources for the RR
potentials of the theory. In non-compact models some of the twisted RR
fields can propagate on non-compact directions and carry the RR-flux off
to infinity. The models are consistent without imposing cancellation of the
corresponding twisted tadpoles. In the compact models we are to consider,
the RR-flux cannot escape to infinity but is trapped in the compact space.
Thus additional constraints may appear.

A second difference is that we consider the simultaneous presence of two
types of D-branes, 9-branes and 5-branes. This is equivalent by T-duality
to considering  both 3-branes and 7-branes in the context of \cite{lr}.
However, in this reference the emphasis is in the gauge theory on the
world-volume of the D3-branes, while the D7-branes are considered
non-dynamical. This is justified since the D7-branes have more non-compact
dimensions than the D3-branes. In our compact models, however, D7 branes
wrap compact direction and yield truly four-dimensional fields. Thus the
cancellation of anomalies from gauge groups living on the D7-branes lead
to additional constraints, not present in the non-compact models. The
considerations in this and the preceding paragraph explain the difference
between our conclusions and those for non-compact models in \cite{lr}.

The outline of the paper is as follows. In the next section we review some
facts about  compact $D=4$, $N=1$ Type IIB orientifolds and establish the
notation needed for the remaining sections. In section 3.1 we study the
compact orientifolds without even order generators i.e., $Z_3$, $Z_7$ and
$Z_3\times Z_3$ orientifolds. In this case the models have only D9-branes
and tadpole conditions uniquely fix the gauge group. However in the
$Z_3\times Z_3$ case it is shown how tadpole cancellation conditions are
stronger than anomaly cancellation conditions. In section 3.2 we study the
compact orientifolds with even order twists. In this case both D9-branes
and D5-branes are present. The $Z_4$, $Z_8$, $Z_8'$ and $Z_{12}'$
orientifolds with standard orientifold projection are shown to be
necessarily anomalous, in agreement with the results of ref. \cite{afiv}
in which  they were shown to have non-vanishing tadpoles.  We present
other examples (the $Z_{12}$ orientifold) in which tadpole conditions are
explicitly shown to be stronger than anomaly cancellation ones. New
solutions for the tadpole conditions are shown for the $Z_6$ and $Z_{12}$
orientifolds leading to a variety of gauge groups previously overlooked in
the literature. We study the cancellation of $U(1)$ anomalies in section~4.
We find that cancellation of non-Abelian anomalies guarantees the cancellation
of Abelian anomalies. In addition it is shown that in the $D=4$  compact
orientifold case cancellation of non-Abelian anomalies also implies the
cancellation of mixed $U(1)$-gravitational anomalies. Section 5 is left for
some final comments and conclusions. In particular we compare our results
to those found for non-compact orientifolds and discuss the origin of
the non-equivalence of tadpole/anomaly conditions. We also briefly discuss
the equivalent results for $D=6$, $N=1$ compact IIB orientifolds.

\section{$D=4$, $N=1$,  Type IIB  Orientifolds}
\label{basics}

In this section we summarize the basic ingredients \cite
{sagnotti,bs,gp,afiv} and notation needed in the construction of
$D=4,N=1$ orientifold. The reader is referred to \cite{afiv}
for further details.

In a Type IIB orientifold, the toroidally compactified theory is
divided out by the joint action of a discrete symmetry group $G_1$,
(like $Z_N$ or $Z_N\times Z_M$) together
with a world sheet parity operation $\Omega$, exchanging left and right
movers. $\Omega $ action can be accompanied by extra operations thus leading
to generic  orientifold group $G_1+ {\Omega} G_2$ with ${\Omega}h {\Omega}
h' \in G_1$ for $h,h' \in G_2$

In this article we will refer to the cases $G_1=G_2$ and $G_1=Z_N$
or $G_1=Z_N\times Z_M$ and such that $D=4$ $N=1$ theories are obtained, when
the twist $\Omega$ is performed on Type IIB compactified on $T^6/G_1$. The
allowed orbifold groups, acting crystalographically on $T^6$ leading to
$N=1$  unbroken supersymmetry were classified in \cite{dhvw}. The list, with
corresponding twist vector eigenvalues $v=(v_1,v_2,v_3)$ associated to the
$Z_N$ orbifold twist $\theta$ is given in Table~\ref{tzn}.

\begin{table}[htb]
\renewcommand{\arraystretch}{1.25}
\begin{center}
\begin{tabular}{|c|c||c|c||c|c|}
\hline
$Z_3$ & $\frac13(1,1,-2)$ & $Z_6^{\prime}$ & $\frac16(1,-3,2)$ &
$Z_8^{\prime}$ & $\frac18(1,-3,2)$ \\
$Z_4$ & $\frac14(1,1,-2)$ & $Z_7$ & $\frac17(1,2,-3)$ &
$Z_{12}$ & $\frac1{12}(1,-5,4)$ \\
$Z_6$ & $\frac16(1,1,-2)$ & $Z_8$ & $\frac18(1,3,-4)$ &
$Z_{12}^{\prime}$ & $\frac1{12}(1,5,-6)$ \\
\hline
\end{tabular}
\end{center}
\caption{$Z_N$ actions in \Deq4.}
\label{tzn}
\end{table}

Orientifolding closed Type IIB string introduces a Klein-bottle unoriented
world-sheet. Amplitudes on such a surface contain tadpole divergences.
In order to eliminate such unphysical divergences Dp-branes must be
generically introduced. In this way, divergences occurring in the
open string sector cancel  up the closed sector ones and  produce a
consistent theory. Tadpole cancellation is interpreted as  cancellation
of the charge carried by RR form potentials. For $Z_N$, with $N$ odd, only
D9-branes are required. They fill the full space-time and six dimensional
compact space. For $N$ even, D$5_k$-branes, with world-volume filling
space-time and the $k^{th}$ complex plane, may be required. This is so
whenever the orientifold group contains the element $\Omega R_i R_j$, for
$k\neq i,j$. Here $R_i$ ($R_j$) is an order two twist of the $i^{th}$
($j^{th}$) complex plane.

In what follows we  consider cases with only one set of five branes.
$Z_N$ twists in Table \ref{tzn} were  organized in such a way that, for
even $N$, the order two element $R=\th^{N/2}$ inverts the complex planes
$Y_1$ and $Y_2$ and thus the corresponding orientifolds have D$5_3$-branes,
filling space-time and compact dimension $Y_3$.

Open string states are denoted by $|\Psi, ab \rangle $, where $\Psi$ refers
to world-sheet degrees of freedom while the $a,b$ Chan-Paton indices are
associated to the open string endpoints lying on D$p$-branes and D$q$-branes
respectively.

These Chan-Paton labels must be contracted with a hermitian matrix
$\lambda ^{pq} _{ab}$.   The action of an element of the orientifold group
on Chan-Paton factors is achieved by a unitary matrix $\gamma _{g,p}$ such
that $g: \lambda ^{pg} \rightarrow \gamma _{g,p} \lambda ^{pq}
\gamma^{-1}_{g,q}$.  We denote by $\gamma _{k,p}$ the matrix associated to
the $Z_N$ orbifold twist $\theta ^k $ acting on a Dp-brane.

Consistency under group transformations imposes restrictions on the
representations $\gamma _g$. For instance, from $\Omega ^2 =1$ it follows
that
\beq
\g_{\Om,p }=\pm\g^T_{\Om,p }
\label{gomt}
\eeq

Tadpole cancellation imposes further constraints on $\gamma _g$. Since we
are planning to compare such restrictions with those coming from anomaly
cancellation of gauge theories on D5 and D9-brane configurations, we will
not impose the former in what follows, but just consider generic actions
obeying the algebraic consistency conditions. Nevertheless, we will
perform a definite choice of signs in (\ref{gomt}), namely
\beqa
\g_{\Om,9 } & = & \g^T_{\Om,9 } \nonumber \\
\g_{\Om,5 } & =& -\g^T_{\Om,5} \label{gpa}
\eeqa
for $\Om $ acting on 9 and on 5-branes. The first condition is the usual
requirement of global consistency of the ten form potential in Type I
theory. Second equation is in agreement with the Gimon and Polchinski
action, analyzed in \cite{gp}. These constraints lead to  $SO(2N_9)$ and
$USp(2N_5)$ groups in the 99 and 55 open string sectors respectively, where
$2N_9$ $(2N_5)$ is the number of D9(5)-branes (an even number is required
by $\Om $ action we will consider). When $N_9=N_5=16$ such conditions
ensure cancellation of untwisted tadpoles. Notice that consistency under
the action of ${(\Om\theta ^k)}^2 =\theta ^{2k}$ and (\ref{gpa}) lead to
\beq
\g_{k,p}^* =  \g_{\Om,p} \, \g_{k,p} \g_{\Om,p}
\label{famp}
\eeq
for $p=9,5$.

Thus, for a $Z_N$ orbifold twist action, with $N=2P$ ($N=2P+1$) for $N$
even (odd), a  generic matrix  $\gamma _{\theta,p}$  can be written as

\beq
\gamma_{1,p}=({\tilde \gamma_{1,p}},{\tilde \gamma _{1,p}}^{*})
\eeq

where  $* $ denotes complex conjugation. ${\tilde \gamma }$ is a
$N_p\times N_p$ diagonal matrix given by
\beq
{\tilde \gamma }_{1,p}   =
\diag  (\cdots,\alpha^{NV_j}I_{n_j^p},\cdots, \alpha^{NV_ P} I_{n_P^p})
\label{gp}
\eeq
with $\alpha = {\rm e}^{2i\pi /N}$.
$V_j=\frac{j}N$ with $j=0,\dots, P$ corresponds to an action ``with
vector structure '' ($\gamma ^N=1$) while $V_j= \frac{2j-1}{2N}$ with
$j=1,\dots,P$ describes an action
 ``without vector structure'
($\gamma_{1,p} ^N=-1$) \footnote{Following the classification introduced
in \cite{blpssw} for six-dimensional models.}. If we choose matrices
$\g_{\Om,9}$ and $\g_{\Om,5}$
\beq
\g_{\Om,9} = \bmat{cc}0&\id_{N_9}\\ \id_{N_9} & 0 \emat
\quad ; \quad
\g_{\Om,5} = \bmat{cc}0&-i\id_{N_5}\\ i\id_{N_5} & 0 \emat
\label{goms95}
\eeq
then  (\ref{gpa}) and (\ref{famp}) are satisfied.
In what follows we will be mainly concerned with
actions  ``without vector structure ''
 whenever D5-branes are present \footnote{This is the
 case most widely studied for compact
orientifolds although models with vector
structure can also be constructed \cite{lykken,afiv}.}. In those cases, the
Chan-Paton matrices for the orbifold twist break the symplectic factors
down to unitary groups.

For later convenience note that the trace of twist matrix above,
or in general of its $k$-th power $\gamma _{k,p}$, reads

\begin{equation}
\Tr\gamma_{k,p}=\sum_{j=0(1)}^{ P}{2n_j^p \cos(2\pi V_j k)}
\label{trg}
\end{equation}
where the sum starts from 0(1) for the ``with (without) vector structure''
actions. In particular, for $k=0$ we obtain $Tr I_p=\sum {2n_j^p}=2N_p$,
the number of Dp-branes.

Moreover, in order to compare  anomaly cancellation conditions, usually
given in terms of the integers $n_j^p$, with tadpole equations, usually
written in terms of above traces, it is also useful to have the former
expressed in terms of the latter. This is easily achieved by performing a
discrete Fourier transformation.  For instance, for the ``without
vector structure'' case, by multiplying both sides of equation
(\ref{trg}) by $\cos(2k\pi V_j)$ and by summing over $k=0,\dots,N-1$ and
using that $\Tr \g_{N-k,p}=-\Tr \g_{k,p}$, we obtain
\begin{equation}
n_j^p= \frac{1}N[ \Tr\gamma _{0,p} + 2\sum_{k=1}^{ P} \Tr\gamma _{k,p}
\cos({2V_jk\pi })]
\label{njp}
\end{equation}
for $j=1,\dots,P$. A similar expression is valid for the shift "with
vector structure" (N odd) where we  also have the $j=0$ term
\begin{equation}
2n_0^p= \frac{1}N[ Tr\gamma _{0,p} + 2\sum_{k=1}^{ P} Tr\gamma _{k,p}]
\label{nop}
\end{equation}

\medskip

The spectrum associated to the 9 and 5-brane
orientifold configuration is easily obtained by working in
a Cartan-Weyl basis (see \cite{afiv}).

The gauge fields living on the world-volume of a D$p$-brane have associated
Chan-Paton factors $\lambda ^p$ corresponding to the gauge group $G_p$
with $G_9=SO(2N_9)$ and $G_5=Sp(2N_5)$. In Cartan-Weyl  basis such generators
are organized into charged generators $\lambda_a = E_a$, $a=1,\cdots, {\rm
dim}\, G_p - {\rm rank}\, G_p$, and Cartan algebra generators $\lambda_I =
H_I$, $I=1,\cdots, {\rm rank}\, G_p$, such that
\begin{equation}
[H_I, E_a]=\rho_I^aE_a
\label{cw}
\end{equation}
where the (${\rm rank}\, G_p$)-dimensional vector with components $\rho_I^a$
is the root associated to the generator $E_a$.

The matrices  $\gamma_{1,p}$ and its powers represent the action of the
$Z_N$ group on Chan-Paton factors, and they correspond to elements of a
discrete subgroup of the Abelian group spanned by the Cartan generators.
Hence, we can write
\begin{equation}
\gamma _{1,p}= e^{-2i\pi V^p \cdot H }
\label{Vdef}
\eeq

Thus, this equation defines a (${\rm rank}\, G_p$)-dimensional vector $V^p$
with  coordinates corresponding  to the $V_j$'s defined in (\ref{gp})
above. Cartan generators are represented as tensor products of $\sigma_3$
Pauli matrices.  In such a description the massless states are easily
found. Let us consider the case in which all 5-branes sit at the origin.
In the $pp$ sector the gauge group is obtained by
selecting the root vectors satisfying
\begin{equation} \rho^a \cdot V^p= 0 {\rm \, mod \,} {\bf Z}
\label{v9p}
\end{equation}
while  matter states correspond to charged generators with
\begin{equation} \rho^a \cdot V^p= v_i {\rm \, mod \,} {\bf Z}
\label{m9p}
\end{equation}

Recall that root vectors for orthogonal groups are of the  form
$\underline {(\pm1,\pm1,0,\cdots,0)}$ where underlining indicates that all
possible permutations must be considered. In the symplectic case we have
to include, in addition, the long roots $\underline {(\pm2,0,\cdots,0)}$.

In the  $95$ sector the subset of roots of $G_9\times G_5$ of the form
\beq
P _{(95)}= (W_{(9)}; W_{(5)})=
({\underline {\pm 1, 0,  \cdots, 0}};{ \underline {\pm 1, 0, \cdots, 0}})
\label{w95def}
\eeq
must be considered.  Matter states are obtained from

\begin{equation}
P_{(95)} \cdot V ^{(95)}= (s_jv_j +s_kv_k) {\rm \, mod \,} {\bf Z}
\label{95sh}
\end{equation}
with $s_j=s_k=\pm \frac1{2}$, plus (minus) sign corresponding to particles
(antiparticles) and  $V^{95}= (V^9; V^5)$.

\bigskip

\section{Tadpoles versus anomalies in $D=4$, $N=1$ orientifolds}

\subsection{Odd order orientifolds}

In this subsection we center on the study of $D=4$ $N=1$ $Z_N$
orientifolds, with odd $N$. These models are consistent without the
introduction of D5-branes, so only D9-branes are included. The only
compact odd order orientifolds correspond to  $Z_3$, $Z_7$ and $Z_3\times
Z_3$ with vector twists given in Table~\ref{tzn}. However, let us
momentarily be more general and consider also $Z_N$ orbifold actions which
are not necessarily crystallographic. Thus, we focus on twist generators
$\theta$ with eigenvalues $\frac{1}N(t_1,t_2,t_3)$ with
$\sum_{a=1}^{3}{t_{a}}=0$ $ {\rm mod} \  N$,  and $N=2P+1$.

The strategy to study the relation between the anomaly and tadpole
cancellation  conditions will be as follows. First we compute the gauge
group and  massless matter for a general such model. In this step only
algebraic  consistency conditions (group law) are imposed on the
Chan-Paton matrices. Next we find generic conditions for cancellation of
gauge anomalies \footnote{Notice that when one of the gauge factors is
absent, the conditions of anomaly cancellation may be less restrictive than
those we consider \cite{lr}. However, this case is rather particular, and
we find it is more insightful to consider `generic' cancellation of
anomalies, as we do in the present paper.}. Since for $SU(n)$ groups only
fundamental ${\bf n}$ and/or antisymmetric ${\bf a}_n$ representations
(or their conjugates) appear, such conditions will manifest as restriction
on the group ranks in order to ensure that only anomaly-free combinations
are allowed. Finally, these restrictions are compared with tadpole
cancellation equations obtained from type IIB orientifolds.

In agreement
with eq.(\ref{gp}), the generic action of the orbifold twist on 9-branes
can be encoded in the  matrix
\beq
\gamma_{1,9}=({\tilde \gamma_{1,9}},{\tilde \gamma}^{*}_{1,9})
\eeq
with ${\tilde \gamma }$ given by
\beq
{\tilde \gamma }   = \diag  (I_{n_0^9}, \alpha I_{n_1^9},
\cdots, \alpha^{j} I_{n_j^9},\cdots, \alpha^P I_{n_P^9})
\label{g9} \eeq
and  $\alpha = {\rm e}^{2i\pi /N}$. The  associated shift is
\begin{equation}
V^{9 }={\frac{1}{N}} (0,\cdots,0,1\cdots 1,\cdots,j,\cdots,j,\cdots, P,
\cdots,P)
\label{v9}
\end{equation}
from where we can easily read the gauge group to be
\begin{equation}
SO(u_0)\times \prod _{j=1}^P U(u_j)
\label{gg9}
\end{equation}
where we have defined $u_0=2n_0^9$ and $u_j=n_j^9$. For the sake of clarity
let us first consider orbifold twists of the form $\frac{1}N(1,1,-2)$.

The corresponding  massless   spectrum is
\begin{eqnarray}
& & 2[{\ov {\bf a}}_P   + ({\bf u}_0,{\bf u}_1)_{(1)} +\sum _{j=1}^ {P-1}
({\ov {\bf u}_j},{\bf u}_{j+1})_{(-1,1)}]+ \nonumber  \\
& & {\ov {\bf a}}_1  +({\bf u}_0,{\ov {\bf u}}_2 )_{(-1)}
+ \sum _{j=1}^ {P-2}({ {\bf u}_j},{\ov {\bf u}}_{j+2})_{(-1,1)}  +
({\bf u}_{P-1}, {\bf u}_P)_{(1,1)}
\label{oddspec}
\end{eqnarray}
where, inside brackets we have indicated the charge with respect to the
$U(1)$ factor in $U(u_j)$.  The absence of $SU(u_j)$ gauge anomalies
requires
\beqa
& & SU(u_1) \quad
: 2 u_0 +u_3-2u_2-u_1+4=0 \nonumber \\
& & SU(u_j) \quad
:   2u_{j-1}+u_{j+2}-2u_{j+1}- u_{j-2} =0\\
& & SU(u_P) \quad :     3u_{P-1} -u_{P -2}-2u_P+8=0 \nonumber
\label{oddgac1}
\end{eqnarray}
where $j\ne 1, P$.
Actually, by performing the identifications $u_j=u_{-j}$ and $u_{N+k}
=u_k$ these conditions  can be recast  into the unique expression
\beq
SU(u_j) \quad :  2u_{j-1}+u_{j+2}-2u_{j+1}- u_{j-2} =
-4\delta _{j,1}-8{\delta _{j,P}}
\eeq
for $j=1, P$.

Moreover, it is not difficult to generalize these equations to the case of
general odd order orbifolds, generated by a twist with eigenvalues
$\frac{1}N (t_1,t_2,t_3)$, with $\sum t_{a}= 0$ ${\rm mod}$ $ N$. We
obtain
\beq
SU(u_j) \quad :  \sum _{a=1}^{3} (u_{j-t_a}-u_{j+t_a })
= 4\sum_{a=1}^{3}
(\delta _{2j,t_a}-\delta _{2j,-t_a})
\label{gaco}
\eeq
for $j=1,\cdots, P$, and the arguments of the Kronecker deltas are
defined mod N.

Such anomaly cancellation conditions can be reexpressed in terms of traces
of the matrix $\gamma$ by using equations (\ref{njp}) and (\ref{nop}) that,
with the above definition of $u_0$ now read
\begin{equation}
u_j= \frac{1}N[ \Tr\gamma _{0,9} + 2\sum_{k=1}^{ P} \Tr\gamma _{k,9}
\cos(\frac{2kj \pi} N)]
\label{uj}
\end{equation}
for $j=0, \dots,P$

Hence, by replacing in (\ref{gaco}) we obtain
\begin{equation}
\frac{1}N \sum_{k=1}^{ P } \sin( \frac{2\pi kj}N)[\sum _{a=1}^3
\sin(\frac{2\pi kt_a}N ) ] Tr \gamma _{k,9} = \sum _{a=1}^{3}
(\delta_{2j,t_a}-\delta _{2j,-t_a})
\label{antrgo}
\end{equation}

By Fourier transforming the delta functions and using that
\beq
\sum _{a=1}^3
\sin(\frac{2\pi k t_a }N) = -4 \prod _{a=1}^3 \sin(\frac{\pi kt_a}N )
\label{ck}
\eeq
we obtain the general conditions for the absence of gauge anomalies
\begin{equation}
\prod _{a=1}^3 \sin( \frac{\pi kt_a}N)\,[\, 2\Tr \gamma _{2k,9} \prod_{a=1}^3
\cos(\frac{\pi kt_a}M) -1 \, ] =0
\label{antrgos}
\end{equation}

Notice that even if $Tr\gamma _{0,9}$ appears in (\ref{uj}), it is not present
in these conditions, there is no dependence on the total number of 9-branes.
 This was expected, given the relation of these gauge
theories to systems of D3 branes at non-compact singularities, to be
discussed below.

For the compact $Z_3$ and $Z_7$  cases these are the well known twisted
tadpole cancellation conditions $Tr \gamma _{1,9}=-4$ and $Tr \gamma
_{2}=4$ respectively \cite{ang,kak3,afiv} . Such
conditions on the traces or,
equivalently, equations (\ref{gaco}), completely fix the values of
$u_j$'s, and thus lead  to the  unique solutions found in the literature,
when the number of 9-branes is 32.

In order to interpret several results in this paper, it will be useful to
relate our models to systems of D3 branes at orientifold singularities
\cite{lykken,kak1,kak2,iru}.
The basic observation is that the gauge theory we have described in the
case at hand can be realized by placing D3 branes at a non-compact $Z_N$
orientifold singularity \cite{iru}. This can be understood by T-dualizing
along the three compact directions, which transforms the D9-branes into
D3-branes sitting at one of the fixed points, and then taking a
decompactification limit. Thus, the conditions (\ref{antrgos}) ensure the
cancellation of anomalies on the D3 brane world-volume. This explains why
$\Tr \gamma_0$ (which in this case is the total number of D3 branes) is
unconstrained: there is an infinite family of non-anomalous theories,
parametrized by the number of D3 branes at the singularity. Actually, in
the non-compact case, the conditions above are exactly equivalent to the
tadpole cancellation conditions, in analogy with the result in \cite{lr}.
We have just seen that in the  particular case of $Z_N$ ($N$ odd)
orientifolds,  compact models also have this property.

\medskip

The same conclusion, however, does not follow for other types of
orientifold, as we show below. To this end, let us consider for instance
orientifolds $Z_{N_1}\times Z_{N_2}$, with $N_1$, $N_2$ odd. As in the
above models, these theories are consistent without the addition of
D5-branes. The analysis of the relation between anomaly cancellation and
tadpole cancellation conditions can be studied following the strategy used
above, so we will be more sketchy, leaving the details for specific
examples. The first step is to compute the spectrum on the D9-brane sector
for a general $Z_{N_1}\times Z_{N_2}$ model, and compute its non-Abelian
anomalies in terms of the ranks of the group factors. Then we perform a
discrete Fourier transform to rewrite them in terms of Chan-Paton matrices
associated to the twists ${1\over N} (t_1,t_2,t_3)$ in the orientifold
group. The resulting anomaly cancellation conditions have exactly the form
(\ref{antrgos}) (It must be understood that the matrix $\gamma$ in
(\ref{antrgos}) will correspond to a product of powers of $\gamma
_{\theta}$ and $\gamma _{\omega}$ associated to the particular twist ${1
\over N}(t_1,t_2,t_3)$, where $\theta $ and $\omega$ are the $Z_{N_1}$,
$Z_{N_2}$ generators, respectively).

The fact that we obtain the same expression for $Z_N$ and $Z_{N_1}\times
Z_{N_2}$ orientifolds (with odd $N$, $N_1$, $N_2$) is related to the fact
that the twists in a $Z_{N_1}\times  Z_{N_2}$ orientifold and in $Z_N$
orientifolds have the same structure (namely, no order two twist is
contained in the group).

In particular the expression (\ref{antrgos}) holds for the compact
$Z_3\times Z_3$ orientifold
with  $\theta $ and $\omega$ described by the eigenvalues $\frac 13(1,-1,0)$
and $\frac 13 (0,1,-1)$.  It is important to observe that the first
coefficient in (\ref{antrgos}) vanishes when one direction is not affected
by the twist. Therefore, for the case under consideration, only one
constraint corresponding to the twist $\theta\omega ^2$ with eigenvalues
$\frac 13 (1,1,-2)$, thus affecting all three complex directions, is found.
It reads
\beq
Tr \gamma _{\theta } \gamma _{\omega ^2} =-4
\label{z3z3ac}
\eeq

Hence, we find that anomaly cancellation is much less restrictive, in this
case, than tadpole cancellation, which also requires \cite{kak3,zwart} the
set
of equations
\beqa \Tr \g_{\th} &=& \Tr \g_{\om} \ \, = \ \, \Tr \g_{\th \om} \
\, = \ \, 8
\label{otras}
\eeqa
associated to the other twists, to be satisfied. Imposing these additional
conditions, the spectrum of the model is completely fixed. Thus, even if a
whole set of $Z_3\times Z_3$ anomaly free models satisfying (\ref{z3z3ac})
can be constructed, there is, however, a unique solution satisfying all
tadpole cancellation conditions.

Let us consider  this case in more detail. Twists $\theta $ and $\omega$
are represented by the matrices
\beqa
{\tilde \gamma} _{\theta }  & =& \diag  (\id_{w_0},\id_{u_1}, \alpha
\id_{u_2},\alpha \id_{u_3},\alpha \id_{u_4})
\nonumber
\\[0.2ex]
{\tilde \gamma} _{\omega }  & =& \diag  (\id_{w_0},\alpha \id_{u_1},
 \id_{u_2},\id_{u_3}, \alpha ^2 \id_{u_4})
\eeqa
with $\alpha = {\rm e}^{2i\pi/3}$.

The gauge group is $SO(2w_0)\times U(u_1) \dots U(u_4)$ and
the massless spectrum can be easily computed to be
\begin{eqnarray}
& & {\ov {\bf a}}_2   +  ({\ov {\bf u}_1},{\bf u}_3)+
( {\bf u}_1,{\bf u}_4)+( {\ov {\bf u}}_3,{\ov {\bf u}}_4)+
({\bf 2w}_0,{\bf u}_2)\nonumber \\
& & {{\bf a}}_4   +  ({\ov {\bf u}_1}, {\ov {\bf u}_3})+
( {\bf u}_1,{\ov {\bf u}}_2)+( {
{\bf u}}_2,{{\bf u}}_3)+ ({\bf 2w}_0,{\ov {\bf u}}_4)\\
& &  {\bf a}_1   +  ({\ov {\bf u}_2},{\bf u}_4)+ ( {\bf u}_3,{\ov {\bf u}}
_4)+(  {\bf u}_2,{\ov {\bf u}}_3)+ ({\bf 2w}_0,{\ov {\bf u}}_1) \nonumber
\label{z3z3spec}
\end{eqnarray}

If we only impose the anomaly cancellation condition
\beq
\Tr \gamma _{\theta} \gamma _{\omega ^2} =2(w_0+u_3)-u_1-u_2-u_4=-4
\label{z3z3acn}
\eeq
many  anomaly-free spectra can be obtained. Indeed, as discussed above,
these models can be related to systems of D3-branes at $Z_{3}\times
Z_{3}$ singularities. For these {\it non-compact} constructions, all
these models are consistent. In the non-compact case the reason why the
tadpole conditions (\ref{otras}) need not be imposed is clear \cite{lr}.
Those conditions correspond to twists which leave one complex plane
unrotated. Thus the RR flux can escape to infinity through those planes
and one does not need to impose the corresponding tadpole cancellation. In
the compact case this is different: even though those planes are unrotated
they are compact and the RR charge is trapped. Thus the twisted RR charge
corresponding to those twists has to be canceled and the conditions
(\ref{otras}) have to be imposed. However, although these conditions are
needed for consistency, they are totally disconnected from the
issue of gauge anomaly cancellation.

\subsection{Even order orientifolds}

For orientifolds including even order twists, there will be in general
both D9-branes and D5-branes present. In this section we consider the
construction of $D=4$, $N=1$ models obtained from orientifold configurations
of  D9 and D5-branes.

As in the previous section, we first consider a general (not necessarily
crystallographic) orbifold twist. In a  first step we  compute the gauge
group and massless spectrum of such models for an arbitrary number of
$2N_9$ ($2N_5$) D9(D5)-branes. Generic conditions for cancellation of
gauge anomalies are then found. Using the discrete Fourier transform,
these restrictions are then compared with tadpole cancellation equations
obtained from type IIB orientifolds. Finally, explicit solutions
to these equations are discussed. In some cases new solutions, other than
those reported in the literature are found. For simplicity we concentrate
on  models with only one set of D5-branes, all of them sitting at the
fixed point at the origin. Other situations are considered through
specific examples.

Let us  consider arbitrary $Z_N$ ($N=2P$) twists with eigenvalues given by
$\frac{1}N (t_1,t_2,t_3)$ with $t_1+t_2+t_3=0$  and $t_3$ an even
integer (thus $t_1,t_2$ are odd).

We concentrate on models without vector structure. Thus, from
eq. (\ref{gp}) we have
\beq
{\tilde \gamma }   =
\diag  (\alpha I_{n_1},\cdots,\alpha^{(2j-1)}I_{n_j},\cdots,
\alpha^{(2P-1)} I_{n_P})
\label{g5}
\eeq
with $\alpha = {\rm e}^{i\pi /N}$ and where we have dropped the
upperindex 5 ($n_j^5=n_j$).

A similar structure is  used in the 9-brane sector just by replacing
the integers  $n_j$'s by $u_j$'s. These matrices correspond to the shifts
\begin{equation}
V^{p }={\frac{1}{2N}} (1,\cdots 1,\cdots,2j-1,\cdots,2j-1,\cdots,2P-1,
\cdots, 2P-1)
\label{vp}
\end{equation}
with $n_j (u_j)$ entries $(2j-1)$ for $p=5 (9)$, with $j=1,\cdots,P$.

The resulting  gauge group is
\begin{equation}
\prod _{j=1}^{P} U(n_j)\times \prod _{j=1}^{P} U(u_j)
\label{gg5}
\end{equation}

The massless states in $55$ sector are

\begin{eqnarray}
 \sum _{a=1}^3 {\sum_{j=1}^{P}}^{\prime}
& & [{({\ov {\bf n}_j},{\bf n}_{j+t_a})_{(-1,1)}}+
{({\ov {\bf n}_j},{\ov {\bf n}}_{[-j-t_a+1]})_{(-1,-1)}}+\nonumber \\
& & ({\bf n}_j,{\ov {\bf n}}_{j-t_a})_{(1,-1)}+
({\bf n}_j,{ {\bf n}}_{[-j+t_a+1]})_{(1,1)}] +\nonumber  \\
{\sum _{j=1}^{P}} & & [{{\bf a}_j}_{(2)}
(\delta _{j,{\frac{t_1+1}2}}+ \delta _{j,{\frac{N+t_1+1}2}})+
{{\ov {\bf a}}_j}_{(-2)}
(\delta _{j,{\frac{-t_1+1}2 }}+ \delta _{j,{\frac{N-t_1+1}2} })+ \nonumber\\
& & (t_1 \rightarrow  t_2 )]
\label{z55}
\end{eqnarray}
while the $59$ sector spectrum is
\begin{eqnarray}
& & \sum _{j=1}^ {P}{({\ov {\bf n}_j},{\bf u}_{j+t})_{(-1,1)}}+
{({\ov {\bf n}_j},{\ov {\bf u}}_{[-j-t+1]})_{(-1,-1)}}+ \nonumber \\
& & {({\bf n}_j,{\ov {\bf u}}_{j-t})_{(1,-1)}}+
 {({\bf n}_j,{ {\bf u}}_{[-j+t+1]})_{(1,1)}}+
({\bf n} \leftrightarrow  {\bf u})
\label{specfn}
\end{eqnarray}

where the indices $j$, $j^{\prime}$ are defined mod $N$. Notice that sums
over bifundamentals with indices $(j,j')$ must be performed such that
$1\le j\le  j'\le  \frac{N}2$.  Also, the primed sum on 55 sector means
that $j=j'$ is not included. In fact, in this case, antisymmetric
representations do appear for $t_a$ odd, i.e. $a=1,2$ for our convention
above. In 59 sector we have defined $t=\frac{t_1+t_2}2$, see (\ref{95sh}).
The spectrum in 99-sector can be obtained from eq.(\ref{z55}) by
replacing $n$'s by $u$'s.

Again, it is straightforward to write down a general condition for the
cancellation of non-Abelian anomalies. It reads
\begin{eqnarray}
SU(n_j) \quad
& : & \sum  _{a=1}^3 [(n_{j-t_a}-n_{j+t_a})+u_{j-t}-u_{j+t} ]=
4 [\delta _{j,{\frac{t_1+1}2}}+ \delta _{j,{\frac{N+t_1+1}2}} \nonumber\\
& & -\delta _{j,{\frac{-t_1+1}2 }}- \delta _{j,{\frac{N-t_1+1}2} }+
(t_1 \rightarrow  t_2) ]
\label{gacg}
\end{eqnarray}
with $j=1,\dots,{P}$ and where  the identifications $n_{j}=n_{-j+1}=n_{j+N}$
are understood. As in the odd twist case, such equivalences are automatically
implemented when $n_j$'s ($u_j$'s) are expressed in  terms of traces
through ec.(\ref{njp}). In fact, by using such equation above we find
that, in order to ensure the absence of gauge anomalies for the 55 group
sector we must have
\begin{equation}
\frac{1}N   \sum_{k=1}^{ P-1 }\sin (2\pi k V_j)[\sin(\frac{t_3\pi k}N ) A_k]=
-\delta _{j,{\frac{t_1+1}2}}-\delta _{j,{\frac{N+t_1+1}2}}
+\delta _{j,{\frac{-t_1+1}2 }}+ \delta _{j,{\frac{N-t_1+1}2} } +
(t_1 \rightarrow  t_2)
\label{antrg}
\end{equation}
with
\begin{equation}
A_k= 4\sin(\frac{t_1\pi k}N) \sin(\frac{t_2\pi k}N )
\Tr\gamma _{k,5,0} +  \Tr\gamma _{k,9}
\label{akg}
\end{equation}
where we have emphasized, by adding a $0$ subscript, that D5-branes are
located at the origin. Again $\delta$ functions appear from the
contribution of antisymmetric representations.
Notice that the sum is actually up to $P-1$ (and not up to $P$). This is due
to the fact that, from our definition (\ref{g5}) of the twist matrix ,
in the case without vector structure ,
\begin{equation}
Tr{\gamma_{P,p}}=0
\label{o2t}
\end{equation}

This equation for the order two twist
matrix, which in this case is automatically satisfied by construction,
 appears in  \cite{afiv} (eq.(2.39)) as  necessary
condition for the cancellation of tadpoles

Again, it will be useful to keep in mind that these gauge theories can be
realized on the world-volume of D3-branes at $Z_N$ orientifold
singularities. In this case, D7-branes are also present in the
configuration. In this non-compact context, it is possible to show that
the anomaly cancellation conditions (\ref{antrg}) are exactly equivalent
to the tadpole cancellation conditions. In the following we show that this
property is in general no longer true in the compact models.

The constraints that we have just obtained may be read as a set of $P$
equations (for each value of $j$) with $P-1$ unknowns
$\sin(\frac{t_3\pi k}N ) A_k$. Thus, the system is, in principle, overdetermined and it
could have no solutions at all, unless not all equations are  really
independent. In fact, we will see that some models are not consistent.

Recall that a similar set of equations, obtained from above simply by
exchanging $\gamma _{k,9}$ and $\gamma _{k,5}$ would be required in order
to avoid gauge anomalies in  the 99 sector groups. This $5 \leftrightarrow 9$
symmetry is due to the fact that all fivebranes have been put at the
origin, and it is a manifestation of T (self) duality. Nevertheless we will
see that despite the symmetry of anomaly  equations, in some cases they
allow for solutions which are not symmetric under the exchange $5
\leftrightarrow 9$. The additional tadpole cancellation equation leads to a
fully T duality invariant spectrum.

The above constraints  are valid for an  arbitrary, not necessarily
chrystalographic, even twist action on 5 and 9-branes.
In what follows we analyze the compact cases shown in table 1. The global
constraint $Tr\gamma _{0,9}=Tr\gamma _{0,5}= 32$ must be imposed in such
cases.

\bigskip
\bigskip

{\bf i)  Non consistent models}

For orbifold groups containing a twist with eigenvalues $\frac{1}4(1,1,-2)$,
namely the orbifold actions { $Z_4$, $Z_8$, $Z_8$'} and $Z_{12}$' the above
equations have no solutions. In fact, these  orientifolds were found in
\cite{afiv} to be ill-defined. Difficulties stem from the presence of a
Klein-bottle tadpole proportional to the volume $V_3$ of the third compact
dimension.

Interestingly enough,  such inconsistencies are  recovered here from the point
of view of anomaly cancellation. This can be easily seen in the  $Z_4$
case. In fact,  (\ref{antrg}) leads to the  incompatible equations
$\frac{\sqrt2}2 A_1=8$ for $j=1$ and $\frac{\sqrt2}2 A_1=-8$ for $j=2$ where
$A_1=  2Tr\gamma _{1,5,0} +  Tr\gamma _{1,9}$. The same situation repeats
in the other cases.

We should emphasize that the above comments for inconsistency refer only
to Chan-Paton twists without vector structure. It is in principle possible
to find consistent $D=4$, $N=1$ orientifolds corresponding to these twists
but with CP twists {\it with } vector structure. Indeed we have found
consistent (non-chiral) examples based on $Z_4$ with vector structure.

\bigskip
\bigskip

{\bf ii)  Consistent models}

For all other twists in Table 1 it is possible to  write down a general,
consistent,  solution to eq. (\ref{antrg}). Namely

\begin{equation}
\sin(\frac{t_3\pi k}N )A_k= -4(1+(-1)^k )[\sin(\frac{t_1\pi k}N) +
\sin(\frac{t_2\pi k}N )]
\label{aksol}
\end{equation}
with $k=1,\dots P$.
Thus, for each twist $\theta ^k$ of order $k$ we obtain a condition which
relates $\Tr\gamma _{k,5}$ and  $\Tr\gamma _{k,9}$ as they show up in
tadpole cancellation equations. Whenever  $k$ is such that $\theta ^k $
twist leaves the direction parallel to the D5-brane untouched, then
$\sin(\frac{t_3\pi k}N )=0$ (thus also $\sin(\frac{t_1\pi k}N) +
\sin(\frac{t_2\pi k}N) =0$) and no constraint is present  for the
corresponding $\Tr\gamma _{k,9}$ and $\Tr\gamma _{k,5}$.

On the other hand, since we must have a similar solution for anomaly
cancellation in $99$ sector groups but with $5\rightarrow 9$ in
(\ref{akg}), we will have
\begin{equation}
[ 4\sin(\frac{t_1\pi k}N) \sin(\frac{t_2\pi k}N )-1]
(\Tr\gamma _{k,5,0} -  \Tr\gamma _{k,9})=0
\label{t9t5f}
\end{equation}
for all other twists with $t_3\pi k \ne 0$ $mod$ $N$.
Notice that $4\sin(\frac{t_1\pi k}N )\sin(\frac{t_2\pi k}N )$ is, by the
Lefschetz fixed point theorem, nothing but the number of fixed points of
twist $\theta ^k$ in the complex directions $(Y_1,Y_2)$. Thus, this
equation implies that whenever points other than the origin are kept fixed
then $\Tr\gamma _{k,5,0} =  \Tr\gamma _{k,9}$. Also notice that only even
values of $k$ could lead to non zero values for $A_k$.

Let us consider the different cases in more detail:

\bigskip
\bigskip

${\bf Z_6}$:

 From eq.(\ref{aksol}) we find $A_1=0$ and $A_2=16$, namely
\beqa
& &  \Tr \g_{1,9}  +    \Tr \g_{1,5,0}  =  0 \nonumber\\
& & \Tr \g_{2,9}  +   3\Tr \g_{2,5,0} \ \ = \ \ 16
\label{z6ac}
\eeqa
These are tadpole cancellation conditions found in \cite{kak3,afiv}.
However, in  reference \cite{afiv}  it is found that tadpole
cancellation at other fixed points, other than the origin, imposes an extra
requirement.  Namely \beq
\Tr \g_{2,9}  +   3\Tr \g_{2,5,J} \ \, = \ \, 4 \label{tcz61}
\eeq with $J=1,\cdots, 8$ denoting   the fixed points of $\th^2$ in the
$(Y_1,Y_2)$ planes, other than the origin, where 5-branes may sit. Since
here all branes are at the origin then $\Tr \g_{2,5,J}=0$. We are led to
\beq
\Tr \g_{2,9}  = \ \, 4
\label{tcz62}
\eeq
This is precisely the extra condition (\ref{t9t5f}) that results when
anomaly cancellation for the $99$ sector groups is required.

Thus, we  learn that for the $Z_6$ orientifold, generic absence of anomalies
and of tadpole divergences are equivalent. Let us stress that to arrive at
this conclusion, analysis of tadpole cancellation  at all fixed points
(and not only at the origin) is needed. In fact, it is equation (\ref{tcz62})
which makes tadpole equations to look symmetric under the exchange of 9 and
5-branes (all at the origin) and thus ensures cancellation of 99 sector
anomalies. It appears to us that this point is not sufficiently clear in
the literature. Sometimes tadpole
equations asymmetric between D9 and D5-branes
are written down for  four- or six-dimensional orientifolds. These are
misleading since, as they stand, they would allow for 5-9 asymmetric
solutions which could be anomalous. Taking into account tadpole
cancellations in fixed points away from the origin does in fact impose
D9-D5 symmetric solutions. This is just the right behaviour, given the
fact that putting all 5-branes at the origin is a selfdual configuration
under T-duality.

\medskip

Let us now look for solutions of the above equations. An interesting
feature is that the first row in (\ref{z6ac}) indicates the possibility of
a local cancellation among 9-branes and 5-branes RR charges that may not
vanish independently. This possibility has not been noticed in the
literature, and we will exploit it here to construct new solutions of the
tadpole equations for all 5-branes at origin. The new possibilities,
as we saw in eq.(\ref{t9t5f}), have to do with the fact that only the
origin is fixed under $Z_6$ twist.

In order to study the solutions it is easier to rewrite
the traces in terms of $n_j$'s and $u_j$'s.
Hence, by using (\ref{trg}) we have

\begin{eqnarray}
\frac{1}{\sqrt3}
\Tr \g_{1,5} &  = & n_1-n_3 \nonumber\\
\Tr \g_{2,5}  & = &  n_1-2 n_2+n_3
\end{eqnarray}
and similar equations for $\Tr \g_{k,9} $ in terms of $u_j$'s.
Replacing (\ref{z6ac})-(\ref{tcz62}) and using that there are 32 D5- and
D9-branes we obtain
\begin{eqnarray}
u_1& = & 12-n_1\nonumber \\
u_2& = & n_2= 4 \nonumber \\
u_3& =& n_1
\label{a9}
\end{eqnarray}
Therefore, from (\ref{gg5}), the gauge groups, depending on
the free integer parameter $n_1 \le 12$, are
\begin{equation}
[U(n_1)\times U(4)\times U(12-n_1)]_{55} \times [U(12-n_1)\times
U(4)\times U(n_1) ]_{99}
\label{ggz6}
\end{equation}
The  matter content may easily be obtained from (\ref{z55}) and
(\ref{specfn}) .  Notice that the solution given in \cite{kak3,afiv} has
$n_1=n_3=6$ implying that $\Tr \g_{1,5}=\Tr \g_{1,9}=0 $. Therefore it
corresponds to the case in which 5 and 9 brane contributions cancel
independently.

\bigskip

${\bf Z_6}${\bf '}:

Absence  of anomalies in the 55 sector leads to
\beqa
& & \Tr \g_{1,9}  -  2\Tr \g_{1,5,0} = 0 \nonumber\\
& & \Tr \g_{2,9} = -8
\label{z6'1}
\eeqa
which again  coincide with equations for tadpole cancellation at the
origin. When the anomaly cancellation conditions from the 99 sector
 are  included the full set of constraints
\beqa
& & \Tr \g_{1,9} =\Tr \g_{1,5} = 0 \nonumber\\
& & \Tr \g_{2,9} = \Tr \g_{2,5} = -8
\label{z6'2}
\eeqa
is obtained. It is completely  equivalent to  cancellation of all twisted
tadpoles, including those  at the other three fixed points in the
transverse directions.

One  aspect of this model may appear puzzling. Since the order three  twist
leaves one complex plane unrotated, one would expect, as it happened in
the $Z_3\times Z_3$ case, that the tadpole constraint for $\gamma _{2,9}$
is not  required for anomaly cancellation. It turns out that the presence
of D5-branes makes things different. Indeed, cancellation of gauge
anomalies for 99 sector groups requires eq.(\ref{aksol}) but with 9 and 5
indices exchanged in (\ref{akg}). Explicitly,
\begin{equation}
A_k= 4\sin(\frac{\pi k}6)\sin(\frac{-3\pi k}6 ) \Tr\gamma _{k,9} +
\Tr\gamma _{k,5}
\label{akz'6}
\end{equation}
and thus, in effect, no constraint on $\Tr\gamma _{k,9}$ is found for
$k=2$. Nevertheless $\Tr\gamma _{2,5}=-8$ is still required. The condition
on $\Tr\gamma _{2,9}$ is really obtained from the 55 sector consistency.
A further discussion on this point is presented in section 5.

Unlike $Z_6$, in this case  charges among 9-branes and 5-branes
must cancel independently. This is due, essentially,
to the presence of extra fixed points.
As in the odd orientifold cases, since now the number of conditions
on the traces and the number of unknowns are the
same,  the  solution presented in \cite{zwart,afiv}
(for all 5-branes at the origin) is unique.

It is instructive  to analyze, from the point of view of anomaly cancellation,
the situation  when part of the 32 D5-branes live at other fixed points.
As an example consider the case when five branes are distributed in groups
of $N^5_L$ branes among the four fixed points $L=0,\dots,3$ of the $\theta
$ twist in the first two planes. The structure of the $55 _L$ gauge group
and matter content at point $L$ is exactly the same as that for the origin
considered above. The spectrum at each point is obtained from
(\ref{z55})
and (\ref{specfn}) just by replacing $n \rightarrow n^L$, with the
condition $\sum_{L=0}^3 (n_1^L+n_2^L +n_3^L)=16$. Thus the anomaly constraints
eq.(\ref{z6'1}) must now be imposed at each point.  Namely
\beqa
& & \Tr \g_{1,9} -2\Tr \g_{1,5,L} = 0 \nonumber\\
& & \Tr \g_{2,9} =  -8
\label{z6'L55}
\eeqa
for $L=0,\dots,3$.

However, constraints involving several fixed points are obtained when the
99 group is considered. The $59$ sector (\ref{specfn}) reads
\beq
\sum _{L=0}^3 [
({\ov {\bf u}}_1, {\ov {\bf n}}_1^L)+
({ {\bf u}}_1, {\ov {\bf n}}_2^L)    +
({ {\bf u}}_2, {\ov {\bf n}}_3^L)   +
({{\bf u}}_3, { {\bf n}}_3^L)      +
({\ov {\bf u}}_2, { {\bf n}}_1^L)]
\eeq
and so, a sum of contributions from all fixed points supporting D5-branes
appears. Thus anomaly constraints will read
\beqa
& &\sum_{L=0}^3 \Tr \g_{1,5,L} -2\Tr \g_{1,9} = 0 \\
& & \sum_{L=0}^3 \Tr \g_{2,5,L} =  -8
\label{z6'L99}
\eeqa
Eqs.(\ref{z6'L55}) and (\ref{z6'L99}) reproduce the tadpole cancellation
equations for this general case. Since D5-branes are now distributed among
different fixed points we expect to be able to achieve charge cancellation
in different ways rather than in the unique form found above. For instance,
first row in equations above tells us that now it could still be possible to
achieve charge cancellation among 9 and 5-branes even if they did not vanish
independently. By writing the traces in terms of group ranks (\ref{trg})
and recalling that $u_1+u_2+u_3=16$, $\sum _{L=0}^3 n_1^L+n_2^L+n_3^L=16$
we find the following consistency constraints
\begin{eqnarray}
& & n_1^L-n_3^L  =  u_1-6= 2-u_3 \nonumber \\
& & u_2 =  \sum_{L=0}^3 n_2^L= 8 \nonumber \\
& & \sum_{L=0}^3 n_1^L+n_3^L  = 8
\label{z'6c}
\end{eqnarray}

\bigskip

${\bf Z_{12}}$

In this example tadpole cancellation appears to be stronger than anomaly
cancellation. Indeed, anomaly cancellation in the 55 sector gives
\beqa
A_k & = & \Tr \g_{k,9}  -    \Tr \g_{k,5,0} \ \, =
\ \, 0 \quad ; \quad k=1,2,5
\nonumber \\[0.2ex]
A_4 & = &   \Tr \g_{4,9}  +   3\Tr \g_{4,5,0} \ \, = \ \, 16
\label{z12t3}
\eeqa
which must be supplemented with the extra, independent, constraint from
the equivalent equations in the 99 sector,
\beq
\Tr \g_{4,9} = \Tr \g_{4,5,0} \ \,  = \ \, 4
\eeq

Notice, however, that there is no constraint for $A_3=\Tr \g_{3,9}  +
2\Tr \g_{3,5,0}  \;$, due to the appearance of the vanishing factor
$\sin(\frac{4\pi k}{12})$  (for $k=3$) associated to the fourth order twist
$\frac{1}4(1,-1,0) $ which leaves the third plane invariant.

Thus, we expect the model will have additional tadpole conditions, not
needed to ensure anomaly cancellation. This additional constraint is the
tadpole condition (2.45) of \cite{afiv}
\beq
\Tr \g_{3,9}  +  2\Tr \g_{3,5,L} = 0,
\label{z12pt2}
\eeq
where $L$ refers to the fixed points of $\th^3$ in the first and second
complex coordinates. In our specific case with all D5 branes at the
origin, the condition reads
\beq
\Tr \g_{3,9} = \Tr \g_{3,5,0} = 0
\label{tcup}
\eeq
These equations constrain the model beyond mere anomaly cancellation. This
resembles much what happened with $Z_3\times Z_3$.

Again, due to the fact that only
the origin is fixed under $Z_{12}$ there are more solutions to the tadpole
equations than those in the literature.

Let us be more explicit.
The generic gauge group is $\prod _{i=1}^6 U(n_i)$ while the 55+59
 massless spectra is given  by (\ref{z55}-\ref{specfn})
\begin{eqnarray}
{\bf a}_1  + {\overline {\bf a}_6} +({\ov {\bf n}_1},{\bf n}_2)+ ({\ov {\bf n}_2},{\bf n}_3)+
({\ov {\bf n}_3},{\bf n}_4)+({\ov {\bf n}_4},{\bf n}_5)+({\ov {\bf n}_5},{\bf n}_6)
\nonumber\\
{\overline {\bf a}_3} +  {\bf a}_4+ ({\ov {\bf n}_1},\ov {\bf n}_5)+({\ov {\bf n}_2},\ov {\bf n}_4)
+({ {\bf n}_2},{\bf n}_6)+( {\bf n}_3,{\bf n}_5) +( {\bf n}_1;\ov {\bf n}_6)+ \nonumber \\
({\bf n}_1,\ov {\bf n}_4)+({\bf n}_2,{\bf n}_3)+({ \ov {\bf n}_1},{\bf n}_5)+ ({ \ov {\bf n}_2},{\bf n}_6)+
({ \ov {\bf n}_3},{ \ov {\bf n}_6}) \nonumber \\
 (\ov {\bf u}_1;\ov {\bf n}_2)+ ( {\bf u}_1;\ov {\bf n}_3)+( {\bf u}_2;\ov {\bf n}_4)+({\bf u}_3;\ov {\bf n}_5)+
({\bf u}_4;\ov {\bf n}_6)+( {\bf u}_5;\ov {\bf n}_6) +u\rightarrow n
\label{z12spec}
\end{eqnarray}

By rewriting the traces above in terms of the group ranks we find that
anomaly cancellation equations (\ref{z12t3}) constrain such ranks to
satisfy
\begin{eqnarray}
& & n_1-n_4= u_1-u_4 \nonumber        \\
& & n_1+n_3=u_1+u_3 \nonumber\\
& & n_1+n_2 = u_1+u_2 \nonumber\\
& & n_2+n_5=u_2+u_5=4\nonumber\\
& & n_1+n_3+n_4+n_6=u_1+u_3+u_4+u_6=12
\label{a8}
\end{eqnarray}

Thus we see that all $u_j$'s may be expressed in terms of 55 integers,
apart from $u_1$, which is a free integer parameter.

It is interesting to notice that if only eq. (\ref{z12t3}) are used then
anomaly free (but not tadpole-free !) models, which have an asymmetric content in the $55$ and
$99$ sector are possible.

For instance, a family of models obeying such constraint but not
(\ref{tcup}) is obtained by choosing
\begin{eqnarray}
n_1& = &n_4=n_5=4\\
 n_2 &= & 0, \ \ n_3=k, \ \ n_6=4-k\\
u_1&=& u_4=u_5=0            \\
 u_2 & = & 4, \ \ u_3=k+4, \ \ u_6=8-k
\label{z12asim}
\end{eqnarray}
with $k\le 4$.
The corresponding gauge groups are
\begin{equation}
[U(4) \times U(k+4) \times U(8-k)]_{99}
\times [U(k) \times U(4-k) \times U(4)^3]_{55}
\end{equation}

These models are chiral and anomaly free, but are not symmetrical with respect
to the exchange between D9 and D5 branes, and so violate (\ref{tcup}).

In fact, Eq. (\ref{tcup}) reads $n_1-n_2+n_4=u_1 -u_2+u_4=4$ and thus
imposes $u_j=n_j$ for all $j=1,\dots, 6$. With this extra condition the
resulting gauge group  is now
\begin{equation}
[ U(n_1) \times U(n_2) \times U(n_3) \times  U(4+n_2-n_1) \times U(4-n_2) \times
U(8-n_2-n_3)]_{55} \times [ same ]_{99} \label{ggz12}
\end{equation}
symmetric under $5\leftrightarrow 9$.
These are the solutions which obey all tadpole cancellation conditions.
For $n_1=n_3=3$ and $n_2=2$ one recovers  the solution given in
\cite{afiv}.

\section{Anomalous $U(1)$'s }
It is known that orientifold compactifications lead to spectra which usually
contain several Abelian factors with non vanishing triangle anomalies.
In \cite{iru} a generalized Green-Schwarz mechanism ensuring cancellation of
such terms was found \footnote{See \cite{sagnanom,blpssw,intri}
 for the analogous mechanism in six dimensions.}. It involves RR
scalars coming from the twisted closed sector of the string.
In this section we would like to indicate how $U(1)$ anomalies actually
cancel in the models we have considered above, whenever generic
cancellation of non Abelian anomalies is ensured.

\subsection{Mixed non-Abelian anomalies}

For concreteness, let us consider the case of even orientifold models. The
arguments below are also valid for odd orientifolds. In this subsection we
center on the mixed anomaly of the $U(1)$ factor in $U(n_i)$ in the
Dp-brane sector with $SU(n_j)$ non-Abelian group in the Dq-brane sector,
denoted $T^{pq}_{ij}$.

Following \cite{iru} this triangle anomaly will be canceled if
$T^{pq}_{ij}+A^{pq}_{ij}=0$, where $A^{pq}_{ij}$ is the Green-Schwarz term
given by
\beq
A_{ij}^{pq}\ = \ {1\over N}\ \ \sum_{k=1}^{N-1} \ C_k^{pq }(v)\
n_i^p\sin2\pi kV^{p }_i \ \cos2\pi kV^{p}_j
\label{masterorient}
\eeq
Here $k$ runs over twisted $Z_N$ sectors, $p$, $q$ run over 5,9 (meaning
5- or 9-brane origin of the gauge boson) and
\beqa
C_k^{pp } & = & \prod _{a=1}^3 2\sin\pi kv_a \quad {\rm for}\;\; p=q
\nonumber\\
 C_k^{59 } & = & 2\sin\pi kv_3
\label{ckpp59}
\eeqa
In principle, this factorization of $U(1)$ anomalies could lead to new
constraints on the spectrum of the model, beyond those imposed by generic
cancellation of non-Abelian anomalies. In the following, we show this is
{\em not} the case.

Sum over $k$ in (\ref{masterorient}) can be performed explicitly. Consider
the $p=q$ sector. By using (\ref{ck}) and orthogonality of cosines we find
\beq
A_{ij}^{pp}\ = -\frac {n_i}2 \sum_{a=1}^3
[\delta _{i,j+t_a}+\delta _{i,-j+t_a+1}-
\delta _{i,-j-t_a+1}- \delta_{i,j-t_a}]
\label{aijpp}
\eeq
and similarly
\beq
A_{ij}^{59}  =
\frac {n_i}2 [ \delta _{i,j+{\frac{ t_3}2 } }+ \delta_{i,-j+{\frac{t_3}2}+1}-
\delta _{i,-j-{\frac{t_3}2}+1}- \delta_{i,j-{\frac {t_3}2}}]
\label{aij59}
\eeq
where arguments of the Kronecker $\delta$ functions are defined mod $N$.
It is straightforward to check from the spectrum given in (\ref{z55})
and (\ref{specfn}) that this indeed cancels the triangle anomaly whenever
$i\ne j$.

The $i=j$ case is a little bit more involved due to  contributions from
antisymmetric representations.  Recall that the contribution to
the triangle anomaly from the antisymmetric ${{\bf a}_j }_{(2)}$ of
$SU(n)$ is, in our normalization, $2 (\frac{n-2}2)$.  Thus,
the $U(1)_j$-$SU(n_j)$ triangle anomaly reads

\begin{eqnarray}
T_{jj}^{pp}\ & = & \frac {n_j}2 \;
[\; \delta _{j,{\frac{t_1+1}2}}+ \delta_{j,{\frac{N+t_1+1}2}}
 -\delta _{j,{\frac{-t_1+1}2 }}- \delta_{j,{\frac{N-t_1+1}2} }+
(t_1 \rightarrow  t_2) \; ]+ \nonumber\\
& & \frac{1}2 \{\;\sum _{a=1}^3 (n_{j-t_a}-n_{j+t_a})+u_{j-t}-u_{j+t} + \\
& & -4 \;[\;\delta _{j,{\frac{t_1+1}2}}+ \delta _{j,{\frac{N+t_1+1}2}}
-\delta_{j,{\frac{-t_1+1}2 }}- \delta _{j,{\frac{N-t_1+1}2} }+
(t_1 \rightarrow  t_2)\;] \;\} \nonumber
\label{tjj}
\end{eqnarray}
The term between curly brackets is nothing but the expression
(\ref{gacg}) for cancellation of general non-Abelian anomalies. Since it
must vanish the remaining contribution cancels the expression
(\ref{aijpp}) for the case $i=j$.

This shows that the condition of cancellation of non-Abelian anomalies
implies the appropriate factorization of $U(1)$ anomalies. It is also easy
to check that cubic $U(1)$ anomalies are similarly canceled by the GS
mechanism without the need of further constraints.

This behaviour could have been guessed, based on our results in Section~3.
There we noticed that the conditions of non-Abelian  anomaly cancellation
are equivalent to the tadpole cancellation
conditions of a system of D3-branes at non-compact orientifold singularities
(possibly in the presence of D7 branes). In this language, the
additional tadpole conditions only arise in compact models, where the RR
flux cannot escape to infinity. Thus the additional conditions are not
required for consistency of the D3 brane system. Since the
cancellation of $U(1)$ anomalies is required for consistency even in the
non-compact case, it cannot depend on any of the additional constraints.
Thus, appropriate factorization of the $U(1)$'s must follow from the
conditions already imposed by cancellation of non-Abelian anomalies, as
we have shown above.

\subsection{Mixed gravitational anomalies}

We can use a similar computation to show that the cancellation of mixed
$U(1)$-gravitational anomalies does not impose further constraints. This
may appear puzzling at first sight, since gravitational anomalies are
relevant only for compact models. One could expect their cancellation
would involve some (or even all) of the additional tadpole cancellation
conditions. So let us give an intuitive argument (of partial validity) to
understand this fact before entering the detailed computation.

Again, we use the realization of these gauge theories in terms of
D3-branes at orientifold singularities. Certainly, in the non-compact
limit gravity propagates in all ten dimensions, not just in the
four-dimensions where the gauge theory lives, and so there is no obvious
reason why the mixed $U(1)$-gravitational anomalies should cancel.
However, some of these singularities can be embedded in compact Calabi-Yau
spaces (not necessarily toroidal orbifolds). This does not change the
spectrum of the field theory on the D3-branes, since they only feel local
physics, but has the effect that gravitational anomalies become relevant,
and must cancel. Since the mixed $U(1)$-gravitational anomaly receives
contributions only from fields living on the D3 branes, it follows that
gravitational anomalies must cancel for non-compact singularities if they
can be embedded in a global model. Following the argument in the previous
subsection, this shows that, for these singularities, factorization of
gravitational anomalies must be automatic once cancellation of non-Abelian
anomalies is imposed.

Obviously, this argument does not apply to general singularities, since
most singularities cannot be embedded in global contexts. The explicit
computation below, however, shows the conclusion concerning the absence of
new constraints is indeed true for any singularity.

Let us consider the case of even order orientifolds (odd order
orientifolds can be analyzed similarly). Starting from the
spectrum given in section 3.2, the mixed anomaly $T_i^{grav.}$ for the
$i^{th}$
$U(1)$ factor (in the 55 sector) can be shown to be
\beqa
T_i^{grav.} & = & \sum_{a=1}^3  (n_{i-t_a} n_i - n_i\, n_{i+t_a}) - n_i
u_{i+t} + n_i u_{i-t} \nonumber \\
& & - n_i\;[\; \delta_{i,\frac{t_1+1}{2}} + \delta_{i,\frac{N+t_1+1}{2}}
-\delta_{i,\frac{-t_1+1}{2}} - \delta_{i,\frac{N-t_1+1}{2}} +
(t_1 \to t_2) \;]
\label{angra1}
\eeqa
The gravitational anomaly for $U(1)$ from the 99 sector has the same
structure, with the replacement $n_i \leftrightarrow u_i$.
For future convenience, let us note that by using the condition for the
cancellation of non-Abelian anomalies (\ref{gacg}), we can rewrite
(\ref{angra1}) as
\beqa
T_i^{grav.} & = & 3n_i \;[\; \delta_{i,\frac{t_1+1}{2}} +
\delta_{i,\frac{N+t_1+1}{2}}
- \delta_{i,\frac{-t_1+1}{2}} - \delta_{i,\frac{N-t_1+1}{2}} +
(t_1 \to t_2) \;]
\label{angra2}
\eeqa

In \cite{iru} it was proposed that the GS contribution canceling this
anomaly has the form
\beqa
{A_i^{grav}} & = & \frac 34 \frac 1N \sum_{k=1}^N \left[ C_k^{55}(v)\,
\Tr \gamma_{k,5}  + C_k^{59}(v)\,  \Tr \gamma_{k,9} \right] n_i
\sin 2\pi k V_i
\label{irugrav}
\eeqa
This can be Fourier-transformed in a by now familiar way, and be
expressed as
\beqa
{A_i^{grav}} & = & -\frac 38 n_i [\; \sum_{a=1}^3 (n_{i-t_a} + n_{-i+1+t_a}
- n_{i+t_a} -n_{-i+1-t_a}) + \nonumber \\
& & (u_{i-t} + u_{-i+1+t} - u_{i+t} + u_{-i+1-t} ) \;]
\eeqa

Recalling the relations $n_i=n_{-i+1}$, $u_i=u_{-i+1}$, this reads
\beqa
{A_i ^{grav.}} & = & -\frac 34 n_i \left[ \;\sum_{a=1}^3 (n_{i-t_a} -
n_{i+t_a}) + (u_{i-t} - u_{i+t}) \right]
\eeqa
which, after imposing the non-Abelian anomaly cancellation conditions
(\ref{gacg}), precisely cancels (\ref{angra2}).

\medskip

As an explicit example of the above discussion consider the $Z_{12}$
orientifold. The GS factors eq.(\ref{irugrav}) are
$A_1 ^{grav}=-3n_1$, $A_3^{grav}=3n_3$, $A_4^{grav}=-3n_4$ and $A_6^{grav}=3n_6$
while $A_2^{grav}=A_5^{grav}=0$.
The mixed gravitational anomalies $T_i$
are easily computed from eq.(\ref{z12spec}).
For instance for the first and second  $U(1)$ factors in the
55 sector we obtain
\begin{eqnarray}
&T_1^{grav.} & = n_1[n_1-n_2+ n_4-2n_5+n_6+u_3-u_2-4]+3n_1\\
&T_2^{grav.} & = n_2[n_1+n_3- n_4-n_5-n_6-u_1+u_4]
\end{eqnarray}
If  non-Abelian anomaly cancellation equations (\ref{z12asim}) are used then
terms inside brackets vanish and the correct contributions to be canceled
by  GS terms are obtained. The other cases proceed in the same way.

\section{ Conclusions and remarks}

We have studied the relationship between cancellation of tadpoles and
anomalies in compact Type IIB $D=4$ vacua. We have found that only a
subset of the tadpole conditions are in general needed in order to get
anomaly cancellations.

It is worth discussing what characterizes the twisted tadpoles whose
cancellation is not required in order to get anomaly cancellation. In
order to do that it is useful to do a T-duality transformation along  the
three complex compact dimensions. Now we have 3-branes and 7-branes
(with their world-volume including the first two complex compact planes)
instead of 9-branes and 5-branes. We can now decompactify and consider the
$D=4$, $N=1$ field theory living in the intersection of the 3-branes and
7-branes. Now, in this non-compact orientifold one has to impose the
tadpole conditions corresponding to a given twist {\it only if} the flux
of the RR charge originating at the fixed point cannot escape to infinity.
For example, as we discussed in section 3, in the $Z_3\times Z_3$
orientifold one has to impose the tadpole condition only for the twist
$v=\frac 13(1,1,-2)$, because for twists leaving one fixed complex plane
like e.g., $v=\frac 13(1,-1,0)$ the RR flux can escape to infinity. In the
$Z_{12}$ case, which is another case in which a particular tadpole equation
(\ref{tcup}) is not needed to get anomaly cancellation, something similar
happens. In this case the twist is of the form $v=\frac 14(1,-1,0)$ and
the RR flux can escape to infinity (in the non-compact case) through the
third complex plane which is transverse to the D7-brane worldvolume. Now,
coming back to the {\it compact} orientifold case, since the massless
charged spectrum from open strings is the same for both compact and
non-compact cases, the models are still going to be anomaly-free. Thus the
extra tadpole equations needed in the compact case will not be related
to anomaly cancellation.

The $Z_6'$ orientifold example shows again how things go. In this case
one would be tempted to say that tadpoles associated to the twist
$2v=1/3(1,0,-1)$ should not be needed for anomaly cancellation, since the
RR-flux could escape through the second complex plane. But that is not the
case because the plane which is transverse to the D7-branes is the third
one, which is rotated, no flux can escape through it.

 From the $D=4$ effective field theory point of view a natural question
emerges. If there are tadpole constraints which are not needed for gauge
anomaly cancellation, what is the low-energy symmetry which is guaranteed
by them? We do not have a definite answer to that question but it seems
sensible to believe that some other type of symmetries are guaranteed by
them. In particular all these models have sigma-model $U(1)$ symmetries
as well as discrete gauge symmetries  and   cancellation of their anomalies could
require the extra constraints \cite{iru2}.

Let us finally comment on the $D=6$ case. In $D=6$, $N=1$ models, anomaly
and tadpole cancellation conditions are totally equivalent since there are
no twists with unrotated complex planes and no possibility for the RR flux
to escape. Indeed, an explicit analysis along the lines we described for
the $D=4$ case gives rise to this equivalence. Concerning new solutions to
the tadpole cancellation conditions, it is easy to see that the $Z_2$,
$Z_3$ and $Z_4$ $D=6$, $N=1$ models of \cite{gj,dp2} have as unique solutions
the ones given in those references. However the $Z_6$ orientifold admits
more solutions than those reported there. The reason for this, as it
happened in the $D=4$ case is that the $Z_6$ twist (unlike the other three
cases) has only one fixed point, the one at the origin. Thus the RR flux
due to D9-branes can partially cancel that from D5-branes and lead to new
solutions. The general group one can obtain is the same as that in
(\ref{ggz6}).

\bigskip

\bigskip

\bigskip

\centerline{\bf Acknowledgements}
We are grateful to  A. Font, R.~Rabad\'an and G. Violero  for useful
discussions. G.A. thanks Departamento de F\'{\i}sica Teorica at UAM
for hospitality and financial support. A.~M.~U. is grateful to
M.~Gonz\'alez for encouragement and support. L.E.I. thank CICYT
(Spain) and the European Commission (grant ERBFMRX-CT96-0045)
 for financial support. The work of A.~M.~U. is supported by
the
Ram\'on Areces Foundation (Spain).

\end{document}